\documentstyle[preprint,aps]{revtex}
\newcommand{\be}{\begin{equation}} \newcommand{\ee}{\end{equation}}
\newcommand{\bea}{\begin{eqnarray}}\newcommand{\eea}{\end{eqnarray}}
\def\lb{\label}
\begin{document}
\draft
\title
{\LARGE Colour-singlet Strangelets at Finite Temperature}
\author{M. G. Mustafa$^{ \ *}$ and  A. Ansari$^{ \ * }$}
\address
{Institute of Physics, Bhubaneswar-751005, India.}
\date{\today}
\maketitle
\begin{abstract}
\noindent Considering massless $u$ and $d$ quarks, and massive
(150 MeV) $s$ quarks in a bag with the bag pressure constant
$B^{1/4} = 145$ MeV, a colour-singlet grand canonical partition
function is constructed for temperatures $T = 1-30$ MeV.
Then the stability of finite size strangelets is studied
minimizing the free energy as a function of the radius of the bag.
The colour-singlet restriction has several profound effects when
compared to colour unprojected case: (1) Now bulk energy per
baryon is increased by about $250$ MeV making the strange quark matter
unbound. (2) The shell structures are more pronounced (deeper).
(3) Positions of the shell closure are shifted to lower $A$-values,
the first deepest one occuring at $A=2$, famous $H$-particle !
(4) The shell structure at $A=2$ vanishes only at $T\sim 30$
MeV, though for higher $A$-values it happens so at $T\sim 20$ MeV.
\end{abstract}
\narrowtext

 \vfil
 \eject

Witten conjectured in 1984 \cite{ew1} that the strange quark
matter (consisting of roughly equal number of $u$, $d$ and $s$
quarks) might be absolutely stable compared to normal matter in
which iorn is the most stable. The possibility of stable/metastable
droplets (strangelets) was also suggested. A few years later
Greiner and coworkers \cite{cg,cg1} also predicted that during
the phase transition from quark-gluon plasma (formed in relativistic
heavy ion collisions) to hadrons metastable droplets could be formed
with baryon number $A\sim 10-30$. Experimental search for strangelets
is still going on in relativistic heavy ion collisions \cite{ar}.
Recently Madsen \cite{jm1,jm2,jm3} has written several papers
on the stability of strangelets following liquid drop model
as well as shell model. Gilson and Jaffe \cite{gj} have studied
the stability of small strangelets, with $A<100$, considering
independent particle shell model approach with massless $u$ and
$d$ quarks, and massive $s$ quark ($m_s=150$ MeV) confined in a bag.
At zero temperature clear shell structures are found at certain
baryon numbers emplying the possibility of metastable strangelets.

In a very recent paper \cite{am} we have extended the work of Gilson
and Jaffe \cite{gj} to finite temperature. Using the eigen modes of
$u$, $d$ and $s$ quarks in the MIT bag with the bag constant
$B^{1/4} = 145$ MeV we have studied in Ref. [9] the thermodynamic
properties of
small strangelets at finite temperature $T=0.5-30$ MeV. This scheme
of calculation has the advantage that at $T=0.5$MeV the shell
structure of strangelets is as transparent as those of Ref. [8]
at zero temperarure. At the same time the study can be made at
higher temperatures. There we found that the shell structure
melts away at $T\geq 10$ MeV. In this study of ours and as well as
of others,
the strangelets were not constrained to be colour-singlet which
should be natural for a bound (confined) system. In those studies
the statistical description of many-body quantum systems are based
on unrestricted grand canonical ensemble average. One is usually
forced to employ approximate one-body statistical operators with
broken symmetries. Then in order to restore the symmetry one
needs a projected statistics with good quantum numbers, like
angular momentum, particle numbers, parity, colour etc.
In the present study we want to have the correct colour symmetry,
that is, we construct a colour-singlet partition function to
study the stability of finite size strangelets at finite
temperature. For a single baryon we have already made such
studies in the past \cite{am1}. This is done following the
standard projection technique in the $SU(3)$ colour gauge
space \cite{au2}.

The grand canonical partition function  for a system of
quarks, anti-quarks and gluons is given by

\be
{\cal Z}_j(\beta, V) = {\rm Tr} \ \left ( {\hat {\cal P}}_j
\ e^{-\beta \hat H} \right ) \ , \lb{cp1}
\ee

\noindent where $\beta \ = \ {1 / T}$, $\hat H$ is the
Hamiltonian of the system and ${\cal P}_j$ is a projection
operator for a symmetry group $\cal G$ (compact Lie group)
having a unitary representation ${\hat U}(g)$ in a Hilbert
space $\cal H$,
\be
{\hat{\cal P}}_j \ = \ d_j \ \int_{\cal G} d \mu (g) \
\chi^\star_j(g) \ {\hat U}(g) \ , \lb{cp2}
\ee
where $d\mu(g)$ is the normalized Haar measure in ${\cal G}$.
$d_j$ and $\chi_j$ , are, respectively, the dimension and the
character of the irreducible representation $j$ of ${\cal G}$. For
$SU(3)$ colour-singlet configuration $d_j \ = \ 1$ and $\chi_j \
= \ 1$, and the partition function for quark-gluon system becomes
\be
{\cal Z}_C(\beta, V) \ = \ \int_{SU(3)} \ d\mu (g) \ {\rm Tr}
\left ( \ {\hat U}(g) \ e^{-\beta {\hat H}} \right ) \ , \lb{cp3}
\ee
\noindent where
\be
\int_{SU(N_C)} \ d\mu (g) \ = \ {1\over {N_C !}} \Big (
\prod^{N_C-1}_{i=1} \int^{\pi}_{-\pi} {{d\theta_i}\over {2\pi}}
\Big ) \Big [ \prod^{N_C}_{j<k} \big ( 2 \sin {{\theta_j-
\theta_k}\over 2} \big )^2 \Big ]  \ \lb{cp4}
\ee
\noindent with $N_C=3$ for $SU(3)$ colour group. The chemical
potential dependent partition function can be written (more
details can be found in Refs. [10,11]) as
\be
{\cal Z}_C(\beta, V, \mu_q) \ = \  \int_{SU(3)}  d\mu(g)
\ e^{\Theta} \ , \lb{cp7}
\ee
\noindent with
\bea
\Theta \ &=& \ \sum_{\alpha} \  \Big [  \ln \big \{ 1 +
{{\chi}_q(g)} \ e^{- \beta \left (\epsilon^{\alpha}_{q}-
\mu_q \right )} \big \} \ + \  \ln \big \{ 1 +
{\chi}^{\star}_{\bar q}(g) \ e^{- \beta \left (
\epsilon^{\alpha}_{\bar q}+ \mu_q \right )} \big \}
\nonumber \\
& & \ -   \ln \big \{ 1 - {\chi}_{\rm adj}(g) \ e^{-
{\beta \epsilon^{\alpha}_{g}}}  \big \} \Big ] \ ,  \lb{cp8}
\eea
\bea
&{\rm {where}} \ \ \ \ \chi_{q}(g) \ & = \ \sum^3_{i=1}
\ e^{i\theta_i } \ , \nonumber \\
&\chi_{\rm {adj}}(g) \  = \ { | \chi_q(g) |}^2 & \ - \ 1 \ =
\ 2 \ + \ \sum^3_{i<j} \ \cos \ (\theta_i \ - \ \theta_j) ,
\lb{cp10}
\eea
are , respectively, the characters of the group in the fundamental
(quarks) and adjoint (gluons) representations. $\theta_i$  is a
class parameter obeying the constraint $\sum^3_{i = 1} \ \theta_i
\ = \ 0 \ \ ( mod \  2\pi ) $.
$\alpha$ runs over discrete single-particle states of
quarks and gluons confined in a bag with $\epsilon^{\alpha}_q$
representing quark energies and $\epsilon^{\alpha}_g$ the gluon
energies. The upper limit of the summation $\alpha$ depends
upon the temperature considered.

The colour unprojected partition function is obtained by putting
${\hat {\cal P}}_j = 1$ in eq.(\ref{cp1})
\be
{\cal Z}_U(\beta, V) = {\rm Tr} \ \left (  \ e^{-\beta \hat
H} \right ) \ . \lb{up1}
\ee
\noindent Then for a system of quarks, anti-quarks and gluons
\be
{\cal Z}_U(\beta, V, \mu_q) \ = \ e^{\Theta} \ , \lb{up3}
\ee
where
\be
\Theta \ = \ \sum_{\alpha} \  \Big [  \ln \big \{ 1 +  \ e^{-
\left ( \beta \epsilon^{\alpha}_{q}- \mu_q \right )} \big \} \ + \
\ln \big \{ 1 + \ e^{- \left ( \beta \epsilon^{\alpha}_{\bar q}
+ \mu_q \right )} \big \} \  - \ \ln \big \{ 1 -  \ e^{- {\beta
\epsilon^{\alpha}_{g}}} \big \} \Big ] \ .  \lb{up4}
\ee
\noindent It is to be noted that in eq.(\ref{up1}) trace is
performed with equal weight without recourse to the
corresponding colour group having unitary representation.

Unlike in Ref.[9] we do not require strangelets to be charge
neutral. However we ignore the Coulomb corrections because charge
($Z$) is very small for small $A$, $Z << A$. We also exclude
here, like in Ref. [9], the phenomenological zero point energy
term due to great uncertainty in the choice of the value of the
parameter involved.
The baryon number $A$ is fixed by adjusting the quark chemical
potential such that the excess number of $q$ over ${\bar q}$ is
$3A$, $viz.$,
\be
\triangle N_q = N_q  -  N_{\bar q}  =  (N_u  -  N_{\bar u}) +
(N_d  - N_{\bar d}) + (N_s -  N_{\bar s}) = 3 A \ , \label{qn}
\ee
\noindent where
\be
{N_i} \ = \ {T} {\partial \over {\partial \mu_i}}
\left ( \ln{{\cal Z}_{\left (U\atop C \right )}}\right ) \
\ , \label{gnd}
\ee
\noindent with $i=q, \ {\bar q}$ and $\mu_{\bar q} = -\mu_{ q}$.

\noindent The energy, $E$ and free energy, $F$ of the system
can be written as
\be
E (T, R) \ = \ T^2{\partial \over {\partial T}} \left
(\ln{{\cal Z}_{\left (U\atop C \right )}}\right ) \ + \ \mu_q
\triangle N_q \ + \ BV  \ \ , \label{e}
\ee
\be
F ( T, R) \ = \ - T\ln{{\cal Z}_{\left (U\atop C \right )}} \
+ \ \mu_q \triangle N_q \ + \ BV  \ \ , \label{f}
\ee
\noindent where $BV$ is the bag volume energy \cite{dj}.

The pressure generated by the participants gas
\be
P \ = \ - \left ( {\partial \over {\partial V}} F(T,R) \right
)_{T,{\triangle}N_q} \ \ , \label{p}
\ee
\noindent is balanced by the bag pressure constant, $B$
leading to the stability condition of the system. Then finally
the equilibrium energy of the system is given by
\be
E(T,R_0) \ = \ 4B V \ = {16\over 3}\pi B R^3_0 \ \ . \label{ee}
\ee
\noindent Actually $R_0$ represents the equilibrium value of the
radius at which the free energy (\ref{f}) is minimum.

First of all we may remark that at temperatures considered here
the contributions from anti-quarks and gluons are insignificant.
Now we can discuss our results. In Fig. 1a we show the variation
of the baryon energy per particle ($E/A$) as a function of
$A$ for colour unprojected case at various temperatures as
indicated there. In Fig. 1b same is shown for colour projected
(colour-singlet) case. The lowest temperature considered is
$T=1$ MeV as we are mainly interested in comparative studies.
Looking at Figs. 1a and 1b the following features constitute our
main results:

\begin{enumerate}

\item For the colour-singlet case the bulk energy per baryon
has increased to about $1.1$ GeV so that strange
quark matter becomes unbound in bulk for $B^{1/4} =145$ MeV
(for colour unprojected case it is $850$ MeV). The metastable
strangelets could eventually decay into nuclei.
We are not discussing here the various kinds of
decay modes as those are discussed in Ref. [8] by Gilson and
Jaffe.

\item The first shell structure appears at $A=2$ instead of at
$6$ when colour-singlet restriction is not imposed. This is a
very significant change as now at $A=2$ the number of $u$, $d$
and $s$ quarks each is $2$. Otherwise there are three $u$-quarks,
three $d$-quarks and zero $s$ quarks (see Fig. 2). This clearly
corresponds to metastable $H$ particle with mass $\approx 2.5$ GeV.

\item For higher baryon numbers also the shell positions have
shifted to lower $A$-values with more stable shell effects.
These are $A$ = 8, 14, 18, 28 $\cdots$. The positions of shells
are likely to be very important in the range $A \leq 30$ as
experimental search to detect stable/metastable strangelets
are going on in relativistic heavy ion collisions \cite{ar}.

\item In Fig. 1b shell structures for higher $A$-values vanish
at $T\geq 20$ MeV compared to about $10$ MeV in Fig. 1a. Infact
for colour-singlet case at $A=2$ the shell structure disappears
only at $T\sim 30$ MeV whereas that for unprojected case at
$A=6$ it vanishes only at $T\sim 20$ MeV.
\end{enumerate}

In Figs. 2a and 2b are displayed the number of $u$, $d$ and $s$
quarks per baryon as a function of $A$ for colour unprojected
and projected cases, respectively. For the colour-singlet
strangelets the fraction of $s$ quarks is relatively high for
very small sized strangelets, $A < 15$, there being a strong
similiarity at $A=2$ in the colour projected case and $A=6$
in the colour unprojected case. In Figs. 3a and 3b we show the
variation of charge per baryon. In these figures shell closures
are very clearly seen with net charge being zero there.
This also implies that for these metastable strangelets
($E/Z$) ratio is very large, as usually expected \cite{gb,jb}.
The strangelets with closed shells are relatively more stable
and chemically less reactive.

To conclude we have shown that colour-singlet restriction
on strangelets has a profound effect. The prediction of
metastable $H$ particle comes out as a natural consequence
of this. Next we plan to study the effect of quadruploe
shape deformation on the stability of colour-singlet
strangelets.

\vfil
\eject
\begin{figure}
\caption{ Energy per baryon ($E/A$) as a function of baryon
number $A$ for different temperatures ($T=$1, 10, 15, 20 MeV)
with $B^{1/4}=145$ MeV and $s-$quark mass $m_s$=150 MeV:(a)
Colour unprojected case, (b) Colour projected(colour-singlet)
case.}
\end{figure}
\begin{figure}
\caption{Quark-fraction as a function of $A$ for different
temperatures ($T=$1, 15, 30 MeV) with $B^{1/4}=145$ MeV
and $s-$quark mass $m_s$=150 MeV (a) Colour unprojected
case, (b) Colour projected case.}
\end{figure}
\begin{figure}
\caption{Same as Fig. 2 for charge-fraction}
\end{figure}
\vfil
\eject
\end{document}